\documentclass[final]{cvpr}

\usepackage{times}
\usepackage{epsfig}
\usepackage{graphicx}
\usepackage{amsmath}
\usepackage{amssymb}

\usepackage{algorithm}
\usepackage{algorithmic}
\usepackage{bm,amsmath,amssymb}
\usepackage{multirow}
\usepackage{bbding}
\usepackage{lipsum}

\DeclareMathOperator*{\argmin}{argmin}
\DeclareMathOperator*{\argmax}{argmax}

\newcommand{\nv}{\boldsymbol{n}}

\newcommand{\xv}{\boldsymbol{x}}
\newcommand{\yv}{\boldsymbol{y}}

\newcommand{\tx}{\bm{x}}
\newcommand{\ty}{\bm{y}}

\newcommand{\tc}{\bm{c}}
\newcommand{\tr}{\bm{r}}

\newcommand{\tu}{\bm{u}}
\newcommand{\tk}{\bm{k}}

\newcommand{\tC}{\textbf{C}}
\newcommand{\tX}{\textbf{X}}
\newcommand{\tY}{\textbf{Y}}

\newcommand{\tA}{\textbf{A}}

\newcommand{\tK}{\textbf{K}}

\newcommand{\tTheta}{\mathbf{\Theta}}

\newcommand{\ts}{\bm{s}}

\newcommand{\tw}{\bm{w}}

\newcommand{\vtheta}{\bm{\theta}}

\usepackage[pagebackref=true,breaklinks=true,colorlinks,bookmarks=false]{hyperref}

\def\cvprPaperID{6576} 
\def\confYear{CVPR 2021}

\begin{document}

\title{Deep Gaussian Scale Mixture Prior for Spectral Compressive Imaging}

\author{Tao Huang\textsuperscript{1} \quad Weisheng Dong\textsuperscript{1}* \quad Xin Yuan\textsuperscript{2}* \quad Jinjian Wu\textsuperscript{1} \quad Guangming Shi\textsuperscript{1}\\
\textsuperscript{1}School of Artificial Intelligence, Xidian University \quad \textsuperscript{2}Bell Labs\\
{\tt\small thuang\_666@stu.xidian.edu.cn \quad wsdong@mail.xidian.edu.cn \quad xyuan@bell-labs.com} \\ {\tt\small jinjian.wu@mail.xidian.edu.cn \quad gmshi@xidian.edu.cn}
}

\maketitle

\newcommand\blfootnote[1]{%
\begingroup
\renewcommand\thefootnote{}\footnote{#1}%
\addtocounter{footnote}{-1}%
\endgroup
}

\blfootnote{* Corresponding authors.}

\pagestyle{empty}  
\thispagestyle{empty} 

\begin{abstract}

In coded aperture snapshot spectral imaging (CASSI) system, the real-world hyperspectral image (HSI) can be reconstructed from the captured compressive image in a snapshot. Model-based HSI reconstruction methods employed hand-crafted priors to solve the reconstruction problem, but most of which achieved limited success due to the poor representation capability of these hand-crafted priors. Deep learning based methods learning the mappings between the compressive images and the HSIs directly achieved much better results. Yet, it is nontrivial to design a powerful deep network heuristically for achieving satisfied results.
In this paper, we propose a novel HSI reconstruction method based on the Maximum a Posterior (MAP) estimation framework using learned Gaussian Scale Mixture (GSM) prior.
Different from existing GSM models using hand-crafted scale priors (e.g., the Jeffrey's prior), we propose to learn the scale prior through a deep convolutional neural network (DCNN). Furthermore, we also propose to estimate the local means of the GSM models by the DCNN. All the parameters of the MAP estimation algorithm and the DCNN parameters are jointly optimized through end-to-end training.
Extensive experimental results on both synthetic and real datasets demonstrate that the proposed method outperforms existing state-of-the-art methods.
The code  is available at \url{https://see.xidian.edu.cn/faculty/wsdong/Projects/DGSM-SCI.htm}.
\end{abstract}

\begin{figure}[htbh]
\centering
\includegraphics[width=0.95\linewidth]{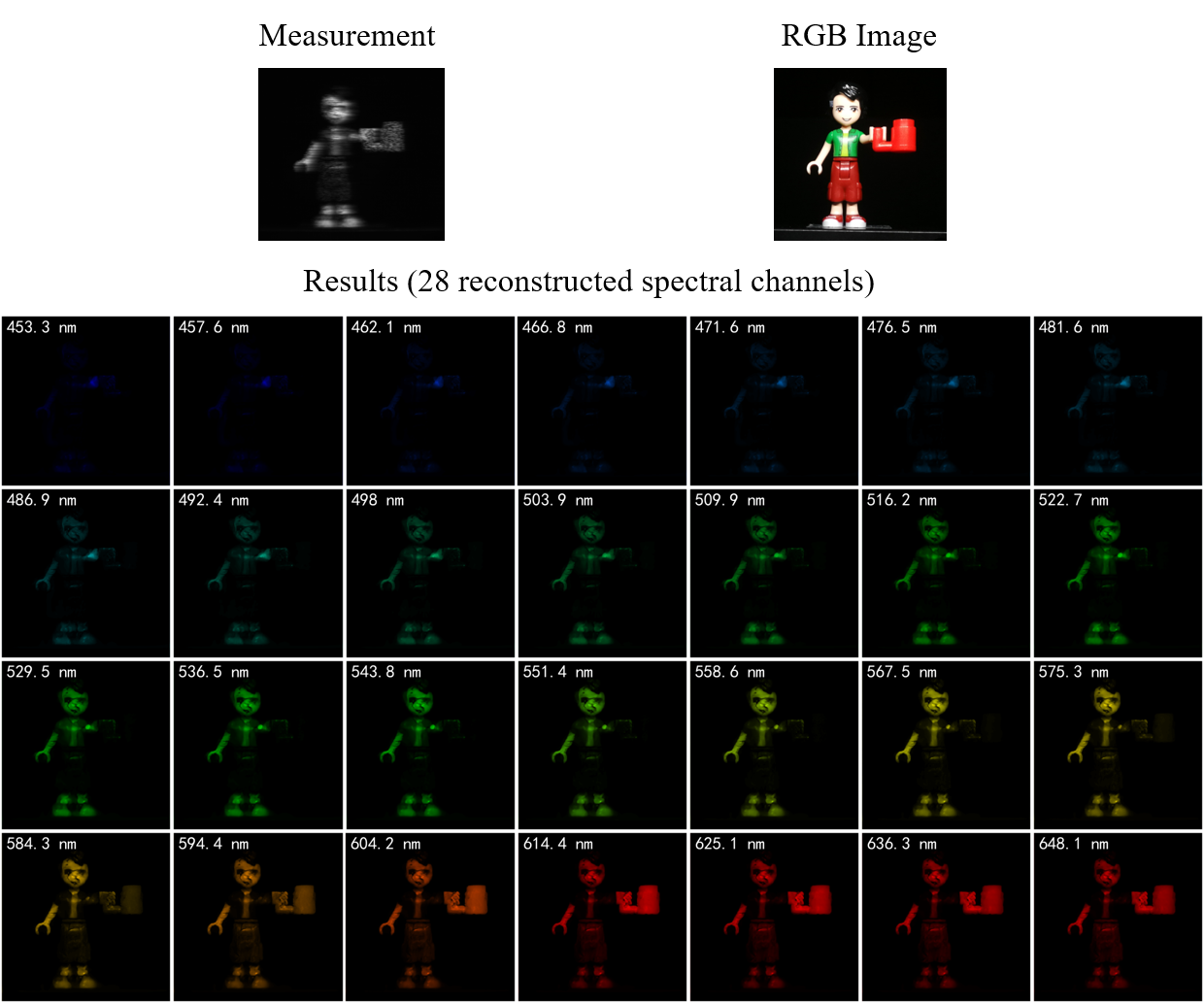}
\caption{A single shot measurement captured by~\cite{Meng2020TSA-Net} and 28 reconstructed spectral channels using our proposed method. }
  \label{fig:Measurement&Result}
\end{figure}

\section{Introduction}
Compared with traditional RGB images, hyperspectral images (HSIs) have more spectral bands and can describe the characteristics of material in the imaged scene more accurately. Relying on its rich spectral information, HSIs are beneficial to many computer vision tasks, \eg, object recognition \cite{uzair2013hyperspectral}, detection \cite{xie2019structure} and tracking \cite{uzkent2016real}. The conventional imaging systems with single 1D or 2D sensor require a long time to scan the scene, failing to capture dynamic objects. Recently, many coded aperture snapshot spectral imaging (CASSI) systems \cite{gehm2007single,Meng2020TSA-Net,Meng2020_OL_SHEM,wagadarikar2008single} have been proposed to capture the 3D HSIs at video rate. CASSI utilizes a physical mask and a disperser to modulate different wavelength signals, and mixes all modulated signals to generate a single 2D compressive image. Then a reconstruction algorithm is employed to reconstruct the 3D HSI from the 2D compressive image. As shown in Fig. \ref{fig:Measurement&Result}, 28 spectral bands have been reconstructed from a 2D compressive image (measurement) captured by a real CASSI system~\cite{Meng2020TSA-Net}.

Therefore, reconstruction algorithms play a pivot role in CASSI. To solve this ill-posed inverse problem, previous model-based methods adopted hand-crafted priors to regularize the reconstruction process. In GAP-TV \cite{yuan2016generalized}, the total variation prior was introduced to solve the HSI reconstruction problem. Based on the assumption that HSIs have sparse representations with respective to some dictionaries, sparse-based methods \cite{kittle2010multiframe, lin2014spatial, wagadarikar2008single} exploited the $\ell_1$ sparsity to regularize the solution. Considering that the pixels of HSIs have strong long-range dependence, non-local based methods \cite{liu2018rank, wang2016adaptive,  zhang2019computational} have also been proposed. However, the model-based methods have to tweak parameters manually, resulting in limited reconstruction quality in addition to the slow reconstruction speed. Inspired by the successes of deep convolutional neural networks (DCNNs) for natural image restoration \cite{lim2017enhanced, zhang2017beyond}, deep learning based HSI reconstruction methods \cite{choi2017high, wang2019hyperspectral, wang2020dnu} have also been proposed.
In \cite{wang2019hyperspectral}, an iterative HSI reconstruction algorithm was unfolded into a DCNN, where two sub-networks were used to exploit the spatial-spectral priors. In \cite{wang2020dnu} the nonlocal self-similarity prior has also been incorporated to further improve the results. In addition to the optimization-inspired methods, DCNN-based methods \cite{Meng2020TSA-Net, miao2019lambda, xiong2017hscnn} that learned the mapping functions between the 2D measurements and the 3D HSIs directly have also been proposed.
$\lambda$-net \cite{miao2019lambda} reconstructed the HSIs from the inputs of 2D measurements and the mask through a two-stage DCNN. TSA-Net \cite{Meng2020TSA-Net} integrated three spatial-spectral self-attention modules in the backbone U-Net \cite{ronneberger2015u} and achieved state-of-the-art results. Although promising HSI reconstruction performance has been achieved, it is non-trivial to design a powerful DCNN heuristically.

Bearing the above concerns in mind, in this paper, we propose an interpretable HSI reconstruction method with learned Gaussian Scale Mixture (GSM) prior.
The contributions of this paper are listed as follows.
\begin{itemize}
    \item Learned GSM models are proposed to exploit the spatial-spectral correlations of HSIs. Unlike the existing GSM models with hand-crafted scale priors (e.g., Jeffrey's prior), we propose to learn the scale prior by a DCNN.
    \item The local means of the GSM models are estimated as a weighted average of the spatial-spectral neighboring pixels. The spatial-spectral similarity weights are also estimated by the DCNN.
    \item The HSI reconstruction problem is formulated as a Maximum a Posteriori (MAP) estimation problem with the learned GSM models. All the parameters in the MAP estimator are jointly optimized in an end-to-end manner.
    \item Extensive experimental results on both synthetic and real datasets show that the proposed method outperforms existing state-of-the-art HSI reconstruction methods.
\end{itemize}

\section{Related Work}
Hereby, we briefly review the conventional model-based HSI reconstruction methods, the recently proposed deep learning-based HSI reconstruction methods and the GSM models for signal modeling.

\subsection{Conventional model-based HSI reconstruction methods}
Reconstructing the 3D HSI from the 2D compressive image is the core of CASSI system and usually with the help of various hand-crafted priors. In \cite{figueiredo2007gradient} gradient projection algorithms were proposed to solve the sparse HSI reconstruction problems. In \cite{lin2014spatial} dictionary learning based sparse regularizers have been employed for HSI reconstruction. In \cite{bioucas2007new, kittle2010multiframe, yuan2016generalized} total variation (TV) regularizers have also been adopted to suppress the noise and artifacts. In \cite{liu2018rank}, the nonlocal self-similarity and the low-rank property of HSIs have been exploited, leading to superior HSI reconstruction performance. The major drawbacks of these model-based methods are that they are time-consuming and need to select the parameters manually.

\subsection{Deep learning-based HSI reconstruction}

Due to the powerful learning ability, deep neural networks treating the HSI reconstruction as a nonlinear mapping problem have achieved much better results than model-based methods. In \cite{xiong2017hscnn} initial estimates of the HSIs were first obtained by the method of \cite{bioucas2007new} and were further refined by a DCNN. $\lambda$-net \cite{miao2019lambda} reconstructed the HSIs through a two-stage procedure, where the HSIs were first initially reconstructed by a Generative Adversarial Network (GAN) with self-attention, followed by a refinement stage for further improvements. In \cite{Meng2020TSA-Net}, DCNN with spatial-spectral self-attention modules was proposed to exploit the spatial-spectral correlation, leading to state-of-the-art performance. Instead of designing the DCNN heuristically, DCNNs based on unfolding optimization-based HSI reconstruction algorithms have also been proposed~\cite{Meng_GAPnet_arxiv2020}. In \cite{wang2019hyperspectral} a HSI reconstruction algorithm with a denoising prior was unfolded into a deep neural network. Since the spatial-spectral prior has not been fully exploited, the method of \cite{wang2019hyperspectral} achieved limited success. To exploit the nonlocal self-similarity of HSIs, the nonlocal sub-network has also been integrated into the deep network proposed in \cite{wang2020dnu}, leading to further improvements.
The other line of work is to apply deep denoiser into the optimization algorithm, leading to a plug-and-play framework~\cite{Zheng20_PRJ_PnP-CASSI}.

\subsection{GSM models for signal modeling}

As a classical probability model, the Gaussian Scale Mixture (GSM) model has been used for various image restoration tasks. In \cite{portilla2003image} the GSM model was utilized to characterize the distributions of the wavelet coefficients for image denoising. In \cite{dong2015image} the GSM model has been proposed to model the sparse codes for simultaneous sparse coding with applications to image restoration. In \cite{ning2020spatial, shi2018robust} the GSM models have also been used to model the moving objects of videos for foreground estimation, achieving state-of-the-art performance. In this paper, we propose to characterize distributions of the HSIs with the GSM models for HSI reconstruction. Different from existing GSM models with manually selected scale priors, we propose to learn both the scale prior and local means of the GSM models with DCNNs. Through end-to-end training, all the parameters are learned jointly.

\begin{figure}[htbh]
\centering
\vspace{-2mm}
\includegraphics[width=0.95\linewidth]{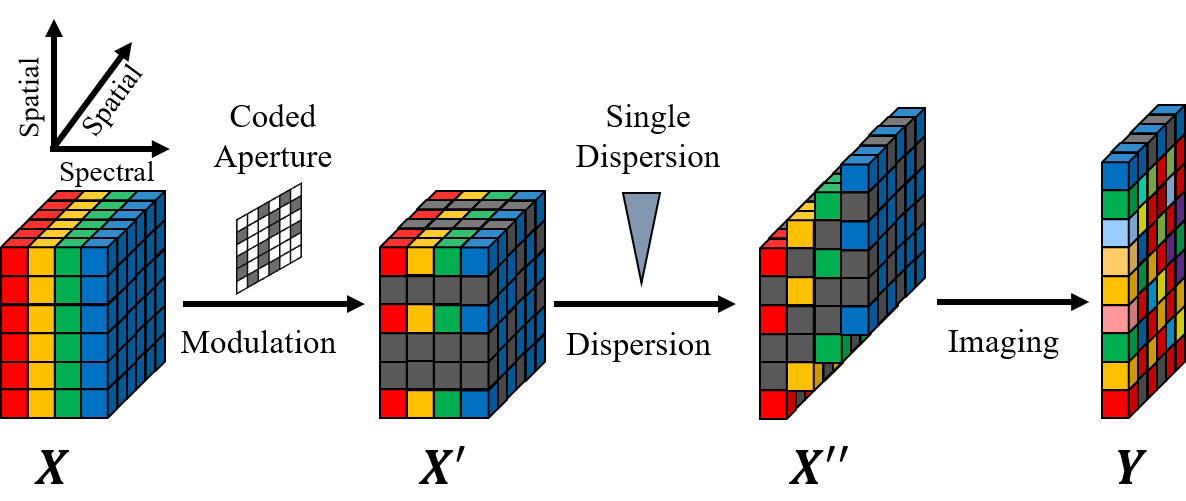}
\vspace{-1mm}
\caption{The imaging schematic of CASSI system.}
\vspace{-2mm}
  \label{fig:CASSI}
\end{figure}

\section{The CASSI Observation Model}
As shown in Fig. \ref{fig:CASSI}, the 3D HSI is encoded into the 2D compressive image by the CASSI system. In the CASSI system, the 3D spectral data cube is first modulated spatially by a coded aperture (i.e., a physical mask). Then, the following dispersive prism disperses each wavelength of the modulated data. A 2D imaging sensor captures the dispersed data and outputs a 2D measurement which mixes the information of all wavelengths.

Let $\tX \in \mathbb{R}^{H \times W \times L}$ denote the 3D spectral data cube and $\tC \in \mathbb{R}^{H \times W}$ denote the  physical mask. The $l^{th}$ wavelength of the modulated image can thus be represented as
\begin{equation}
\tX_{l}^{'} = \tC \odot \tX_{l},
\label{The Modulated Image}
\end{equation}
where $\tX^{'} \in \mathbb{R}^{H \times W \times L}$ is the 3D modulated image and $\odot$ denotes the element-wise product. In CASSI system, the modulated image is dispersed by the dispersive prism. In other words, each channel of the tensor $\tX^{'}$ will be shifted spatially and the shifted tensor $\tX^{''} \in \mathbb{R}^{H \times (W+L-1) \times L}$ can be written as
\begin{equation}
\tX^{''}(r,c,l) = \tX^{'}(r,c+d_{l},l),
\label{The Shifted Image}
\end{equation}
where $d_l$ denotes the shifed distance of the $l^{th}$ channel, $1 \leq r \leq H$, $1 \leq c \leq W$ and $1 \leq l \leq L$. At last, the 2D imaging sensor captures the shifted image into a 2D measurement (by compressing the spectral domain), as
\begin{equation}
\textstyle \tY = \sum_{l=1}^{L} \tX_{l}^{''},
\label{The Compressive Image}
\end{equation}
where $\tY \in \mathbb{R}^{H \times (W+L-1)}$ represents the 2D measurement. As such, the matrix-vector form of Eq. \eqref{The Compressive Image} can be formulated as
\begin{equation}
\ty = \tA \tx,
\label{Observation_Model_matrix-vector}
\end{equation}
where $\tx \in \mathbb{R}^{N}$ and $\ty \in \mathbb{R}^{M}$ denote the vectorized form of $\tX$ and $\tY$ respectively, $N = HWL$ and $M = H(W+L-1)$, and $\tA \in \mathbb{R}^{M \times N}$ denotes the measurement matrix of the CASSI system, implemented by the coded aperture and disperser.
Considering the measurement noise $\nv \in {\mathbb R}^{M}$, the  forward model of CASSI is now
\begin{equation}
    \yv = \tA\xv + \nv.
\end{equation}
The theoretical performance bounds of CASSI have been derived in~\cite{Jalali19TIT_SCI}.

\section{The Proposed Method}
\subsection{GSM models for CASSI}
We formulate the HSI reconstruction as a maximum a posteriori (MAP) estimation problem. Given the observed measurement $\ty$, the desired 3D HSI $\tx$ can be estimated by maximizing the posterior
\begin{equation}
\textstyle  \log p(\tx|\ty) \propto \log p(\ty|\tx) + \log p(\tx),
\label{MAP}
\end{equation}
where $p(\ty|\tx)$ is the likelihood term and $p(\tx)$ is the (to be determined) prior distribution of $\tx$. The likelihood term is generally modeled with a Gaussian function as
\begin{equation}
\textstyle  p(\ty|\tx) = \frac{1}{\sqrt{2\pi}\sigma}\exp\left(-\frac{||\ty - \tA \tx||_{2}^{2}}{2\sigma^{2}}\right).
\label{Likelihood Term}
\end{equation}
For the prior term $p(\tx)$, we propose to characterize each pixel $x_i$ of the HSI with a {\em nonzero-mean} Gaussian distribution of standard deviation $\theta_i$. With a scale prior $p(\theta_i)$ and the assumption that $\theta_i$ and $x_i$ are independent, we can model $\tx$ with the following GSM model
\begin{equation}
\textstyle  p(\tx) = \prod_{i} p(x_i), \quad p(x_i) = \int_{0}^{\infty} p(x_i|\theta_i)p(\theta_i) d\theta_i,
\end{equation}
where $p(x_i|\theta_i)$ is a nonzero-mean Gaussian distribution with variance $\theta_i^2$ and mean $u_i$, i.e.,
\begin{equation}
\textstyle  p(x_i|\theta_i) = \frac{1}{\sqrt{2\pi}\theta_i}\exp(-\frac{(x_i - u_i)^{2}}{2\theta_i^{2}}).  \label{Cond-Gauss}
\end{equation}
With different scale priors, the GSM model can well express many distributions.

Regarding the scale prior $p(\theta_i)$, instead of modeling $p(\theta_i)$ with an exact prior (e.g., the Jeffrey's prior $p(\theta_i)=\frac{1}{\theta_i}$), we introduce a general form as
\begin{equation}
p(\theta_i) \propto \exp(-J(\theta_i)),
\label{scale-prior}
\end{equation}
where the $J(\theta_i)$ is an energy function. Instead of computing an analytical expression of $p(x_i)$ that is often intractable, we propose to jointly estimate $\tx$ and $\bm{\theta}$ by replacing $p(\tx)$ with $p(\tx, \bm{\theta})$ in the MAP estimator. This is
\begin{equation}
\begin{split}
(\tx, \vtheta) &= \textstyle \argmax_{\tx, \vtheta} \log p(\ty|\tx) + \log p(\tx, \vtheta) \\
&= \argmax_{\tx, \vtheta} \log p(\ty|\tx) + \log p(\tx| \vtheta) + \log p(\vtheta).
\end{split}
\label{MAP2}
\end{equation}
By substituting the Gaussian likelihood term of Eq. (\ref{Likelihood Term}) and the prior terms of $p(x_i|\theta_i)$ and $p(\theta_i)$ into the above MAP estimator, we can obtain the following objective function
\begin{align}
\textstyle  (\tx, \vtheta) &= \argmin_{\tx, \vtheta} ||\ty - \tA \tx||_{2}^{2} + \sigma^{2} \sum_{i=1}^{N} \frac{1}{\theta_{i}^{2}}(x_i  - u_i)^{2} \nonumber\\
& \qquad\qquad \textstyle + 2\sigma^{2}\sum_{i=1}^{N}\log\theta_i + 2\sigma^{2} J(\vtheta) \nonumber\\
&= \argmin_{\tx, \vtheta} ||\ty - \tA \tx||_{2}^{2} + \sigma^{2} \sum_{i=1}^{N} \frac{1}{\theta_{i}^{2}}(x_i  - u_i)^{2} \nonumber\\
& \qquad\qquad \textstyle + R(\vtheta),
\label{obj}
\end{align}
where $R(\vtheta)=2\sigma^{2}\sum_{i=1}^{N}\log\theta_i + 2\sigma^{2} J(\vtheta)$. Thereby, the HSI reconstruction problem can be solved by alternating optimizing $\tx$ and $\vtheta$.

For the $\tx$-subproblem, with fixed $\vtheta$, we can solve $\tx$ by solving
\begin{equation}
\textstyle \tx  =  \argmin_{\tx} ||\ty - \tA \tx||_{2}^{2} + \sum_{i=1}^{N} w_i (x_i  - u_i)^{2},
\label{x subproblem}
\end{equation}
where $w_i = \frac{\sigma^{2}}{\theta_{i}^{2}}$ and the mean $u_i$ keeps updating with $\tx$. Inspired by the auto-regressive (AR) model \cite{dong2011image}, we can calculate the weighted average of the local spatial-spectral neighboring pixels as the estimation of the mean $u_i$, \ie,
\begin{equation}
u_{i} = \tk_{i}^{\top} \tx_{i},
\label{auto-regressive model}
\end{equation}
where $\tk_{i} \in \mathbb{R}^{q^{3}}$ denotes the vectorized 3D filter of size $q \times q \times q$ for $x_{i}$ and $\tx_{i} \in \mathbb{R}^{q^{3}}$ represents the local spatial-spectral neighboring pixels of $x_{i}$. For the 3D filters, some existing methods (e.g., the guided filtering \cite{he2012guided, li2013image}, the  nonlocal  means  methods \cite{buades2005non, coupe2008optimized} or the deep learning based method \cite{mildenhall2018burst}) can be used to estimate the spatially-variant filters.

To solve Eq. \eqref{x subproblem}, we employ gradient descent as
\begin{equation}
\tx^{(t+1)} =  \tx^{(t)} - 2 \delta \{ \tA^{\top}(\tA \tx^{(t)} - \ty)+\tw^{(t)} (\tx^{(t)} - \tu^{(t)}) \},
\label{x solution}
\end{equation}
where $\tu^{(t)}=[u_{1}^{t}, \cdots, u_{N}^{t}]^{\top} \in \mathbb{R}^{N}$, $\tw^{(t)}=[w_{1}^{t}, \cdots, w_{N}^{t}]^{\top} \in \mathbb{R}^{N}$ and $\delta$ is the step size.

The $\vtheta$-subproblem can be changed to estimate $\tw$. With fixed $\tx$, $\tw$ can be estimated by
\begin{equation}
\begin{split}
\textstyle \tw  = \argmin_{\tw} \sum_{i=1}^{N} w_i (x_i  - u_i)^{2} + R(\tw).
\end{split}
\label{w subproblem}
\end{equation}
The solution of $\tw$ depends on $R(\tw)$ being used. For some priors, a closed-form solution can be achieved~\cite{ning2020spatial}; for others, iterative algorithms might be used. However, each of them has their pros and cons.
To cope with this challenge, hereby instead of using a manually designed proximal operator, we propose to estimate $\tw^{(t+1)}$ from $\tx^{(t+1)}$ using a DCNN as will be described in the next subsection.


\begin{figure*}[htbh]
\centering
\includegraphics[width=0.85\linewidth,height=0.45\linewidth]{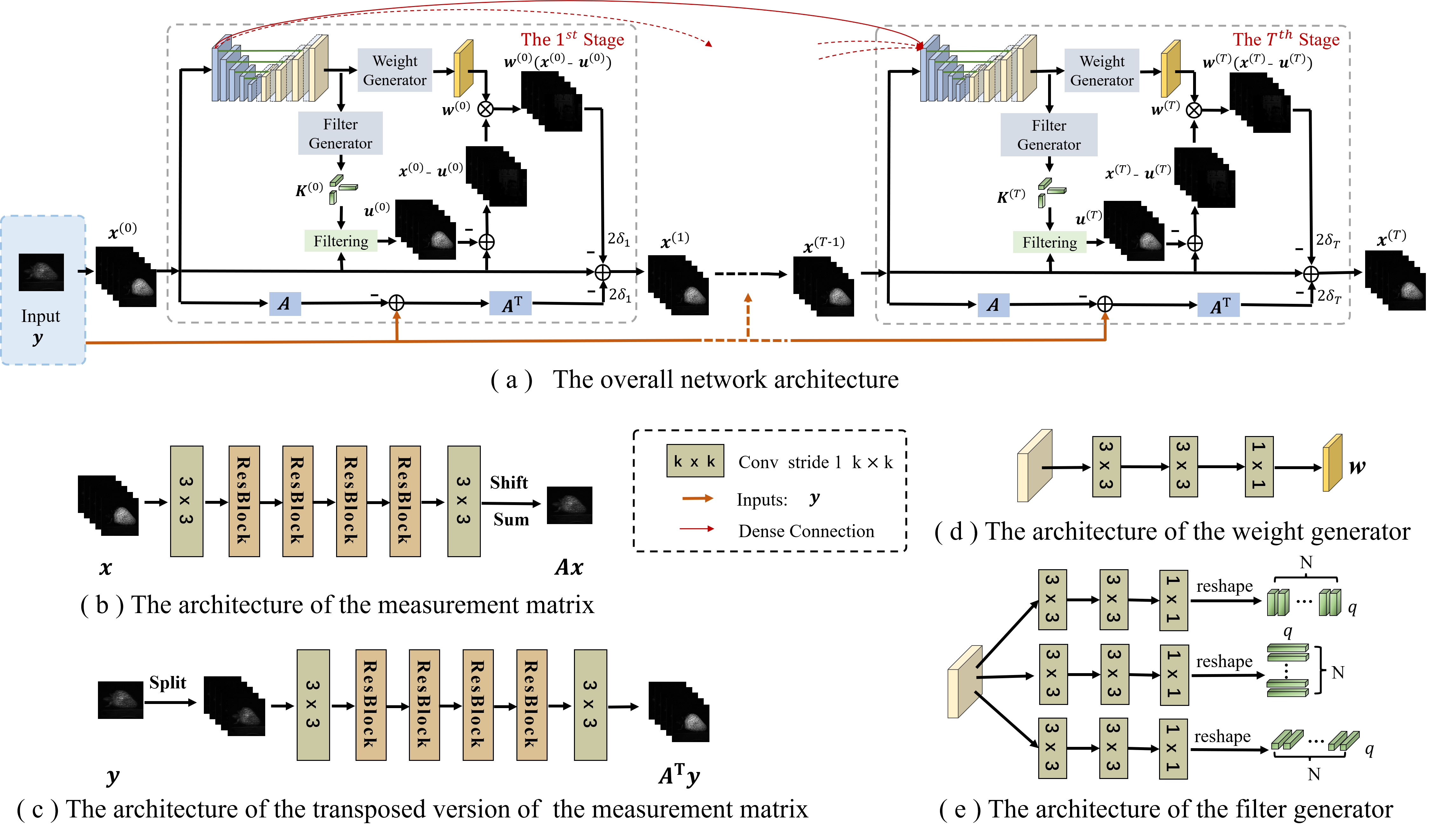}
\caption{Architecture of the proposed network for hyperspectral image reconstruction. The architectures of (a) the overall network, (b) the measurement matrix, (c) the transposed version of the measurement matrix, (d) the weight generator, and (e) the filter generator.}
\vspace{-2mm}
\label{fig:Network}
\end{figure*}

\subsection{Deep GSM for CASSI}
In general,  alternating computing $\tx$ and $\tw$ requires numerous iterations to converge and it is necessary to impose a hand-crafted prior of $p(\vtheta)$. Moreover, all the algorithm parameters and the 3D filters cannot be jointly optimized. To address these issues, we propose to optimize $\tx$ and $\tw$ jointly by a DCNN. For network design purpose, we re-bridge the $\tx$ and $\tw$-subproblems via a united framework
\begin{equation}
\small
\tx^{(t+1)} =  \tx^{(t)} - 2 \delta \{ \tA^{\top}(\tA \tx^{(t)} - \ty)+ {\cal S}(\tx^{(t)}) (\tx^{(t)} - \tu^{(t)}) \},
\label{The united framework}
\end{equation}
where ${\cal S}(\cdot)$ represents the function of the DCNN for estimating $\tw$, \ie, the solution of~\eqref{w subproblem}. As shown in Fig. \ref{fig:Network}(a), we construct the end-to-end network with $T$ stages corresponding to $T$ iterations for iteratively optimizing $\tx$ and $\tw$. The proposed network consists of the following main modules.
\begin{itemize}
    \item The measurement $\ty$ is split into a 3D data cube of size $H \times W \times L$ to initialize $\tx$.
    \item We use two sub-networks to learn the measurement matrix $\tA$ and its transposed version $\tA^{\top}$.
    \item For estimating $\tw$, we develop a lightweight variant of U-Net and a weight generator to learn the function ${\cal S}(\cdot)$.
    \item  Instead of using a manually designed method to learn the 3D filters, we utilize the same lightweight U-Net and a 3D filter generator to generate the spatially-variant filters. According to Eq. \eqref{auto-regressive model}, we filter the current $\tx$ by the generated 3D filters for updating the means $\tu$.
\end{itemize}

\subsection{Network Architecture}

Considering that the real system has large spatial size of the mask and measurements (e.g., the mask and measurements of \cite{Meng2020TSA-Net} are $660 \times 660$ and $660 \times 714$), the network training with explicitly constructed $\tA$ and $\tA^{\top}$ requires a large amount GPU memory and computational complexity. To address this issue, we propose to learn these two operations with two sub-networks. 

\vspace{-2mm}
\paragraph{The modules for learning the measurement matrix $\tA$ and $\tA^{\top}$.} Learning $\tA$ and $\tA^{\top}$ with sub-networks allows one to train them on small patches (e.g., $64 \times 64$ or $96 \times 96$) that can greatly reduce memory consumption and computational complexity. Furthermore, we can train a sub-network to learn multiple masks such that the trained network can work well on multiple imaging systems.
The measurement matrix $\tA$ represents a hybrid operator of modulation, \ie, shifting and summation, which can be implemented by two Conv layers and four ResBlocks followed by shifting and summation operations. As shown in Fig. \ref{fig:Network}(b), $\tx$ is fed into the sub-network to generate modulated feature maps that are further shifted and summed along the spectral dimension to generate the measurements $\ty=\tA \tx$. Each ResBlock \cite{he2016deep} consists of 2 Conv layers with a ReLU nonlinearity function plus a skip connection.
Regarding $\tA^{\top}$, as shown in Fig. \ref{fig:Network}(c), we first slide a $H \times W$ extraction window on the input $\ty$ of size $H \times (W+L-1)$ with the slide step one pixel and split the input into $L$-channel image of size $H \times W$. Then the split sub-images are fed into two Conv layers and four ResBlocks to generate the estimate $\tA^{\top}\ty$.

\begin{figure}[htbp]
\centering
\vspace{-2mm}
\includegraphics[width=0.95\linewidth]{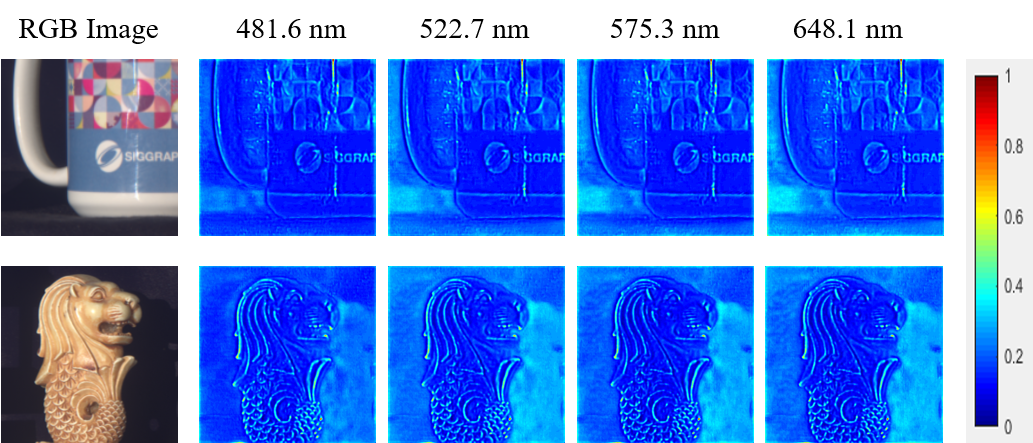}
\caption{The visualization of the regularization parameters $\tw$ estimated in the 4-th stage. Left: the corresponding RGB image; right: the $\tw$ images associated with the four spectral bands. }
\vspace{-4mm}
\label{fig:w}
\end{figure}

\vspace{-2mm}
\paragraph{The module for estimating the regularization parameters $\tw$.} As shown in the Fig. \ref{fig:Network} (a), we propose a lightweight U-Net consisting of five encoding blocks (EBs) and four decoding blocks (DBs) to estimate the weights $\tw^{(t)}$ from the current estimate $\tx^{(t)}$. Each EB and DB contains two Conv layers with ReLU nonlinearity function. The average pooling layer with a stride of 2 is inserted between every two neighboring EBs to downsample the feature maps and a bilinear interpolation layer with a scaling factor 2 is adopted ahead of every DB to increase the spatial resolutions of the feature maps.
We have noticed that the average pooling works better than max pooling in our problem and the bilinear interpolation plays an important role in DBs.
$3\times 3$ Conv filters are used in all the Conv layers. The channel numbers of the output features of the 5 EBs and 4 DBs are set to 32, 64, 64, 128, 128, 128, 64, 64 and 32, respectively. To alleviate the gradient vanishing problem, the feature maps of the first EB are connected to first EB of the U-net of the subsequent stages. The feature maps of the last DB are fed into a weight generator that contains 2 $3\times 3$ Conv layers to generate the weights $\tw$ as shown in Fig. \ref{fig:Network} (d). Some weight maps $\tw$ of two HSIs estimated in the fourth stage are visualized (with normalization) in Fig. \ref{fig:w}. From Fig. \ref{fig:w}, we can see that $\tw$ vary spatially and are consistent with the image edges and textures. Aided by this well-learned $\tw$, the proposed method will pay attentions to the edges and textures.

\vspace{-2mm}
\paragraph{The module for estimating the local means $\tu$.} We estimate the means of GSM models following Eq. \eqref{auto-regressive model}. To estimate the spatial-variant 3D filters, we add a filter generator with the input of the feature maps generated by the U-net, as shown in Fig. \ref{fig:Network}(a). Estimating the spatially adaptive 3D filters has advantages in adapting to local HSI edges and texture structures. However, directly generating these 3D filters will cost a large amount of GPU memory that is unaffordable. To reduce the GPU memory consumption, we propose to factorize each 3D filter into three 1D filters, expressed as
\begin{equation}
\tK_{i} =  \tr_{i} \otimes \tc_{i} \otimes \ts_{i},
\end{equation}
where $\tK_{i}\in\mathbb{R}^{q\times q \times q}$ denotes the 3D filter, $\tr_{i}\in\mathbb{R}^q$, $\tc_{i}\in\mathbb{R}^q$ and $\ts_{i}\in\mathbb{R}^q$ denote the three 1D filters corresponding to the three dimensions, respectively, and $\otimes$ denotes the tensor product. In this way, filtering the local neighbors $\tX_i$ with the 3D filter $\tK_{i}$ can be transformed into convoluting the local neighbors with the three 1D filters along three dimensions in sequence. By factorizing each 3D filter into three 1D filters we can reduce the number of filter coefficients from $N\cdot q^3$ to $3\cdot N\cdot q$, and thus significantly reduce the GPU memory cost and the computational complexity. As shown in Fig. \ref{fig:Network}(e), the filter generator contains three branches to learn the 1D filters, respectively. After generating the filters, we can compute the means of GSM models following Eq. \eqref{auto-regressive model}.


\subsection{Network training}

We jointly learn the network parameters $\tTheta$ through end-to-end training. 
Except the step size $\delta$, all the network parameters of each stage are shared. All the parameters are optimized by minimizing the following loss function
\begin{equation}
\textstyle \hat{\tTheta} = \argmin_{\tTheta} \frac{1}{D} \sum_{d=1}^{D} \|\mathcal{F}(\ty_d; \tTheta) - \tx_d\|_1,
\label{Loss}
\end{equation}
where $D$ denotes the total number of the training samples, $\mathcal{F}(\ty_d; \tTheta)$ represents the output of the proposed network given $d^{th}$ measurement $\ty_{d}$ and the network parameters $\tTheta$, and $\tx_d$ is the ground-truth HSI. The ADAM optimizer \cite{kingma2014adam} with setting $\beta_1 = 0.9$, $\beta_2 = 0.999$ and $\epsilon = 10^{-8}$ is exploited to train the proposed network. We set the learning rate as $10^{-4}$. The parameters of the convolutional layers are initialized by the Xavier initialization \cite{he2015delving}. We implement the proposed method in PyTorch and train the network using a single Nvidia Titan XP GPU. Instead of using the $\ell_2$ norm in the loss function, here we use the $\ell_1$ norm that has been proved to be better in preserving image edges and textures.


\section{Simulation Results}
\subsection{Experimental Setup}

To verify the effectiveness of the proposed HSI reconstruction method for CASSI, we conduct simulations on two public HSI datasets CAVE~\cite{yasuma2010generalized} and KAIST~\cite{choi2017high}. The CAVE dataset consists of 32 HSIs of spatial size $512 \times 512$ with 31 spectral bands. The KAIST dataset has 30 HSIs of spatial size $2704 \times 3376$ also with 31 spectral bands. Similar to TSA-Net~\cite{Meng2020TSA-Net}, we employ the {\em real mask} of size $256 \times 256$ for simulation. Following the procedure in TSA-Net~\cite{Meng2020TSA-Net}, the CAVE dataset is used for network training, and 10 scenes of spatial size $256 \times 256$ from the KAIST dataset are extracted for testing. To be consistent with the wavelength of the real system~\cite{Meng2020TSA-Net}, we unify the wavelength of the training and testing data by spectral interpolation. Thus, the modified training and testing data have 28 spectral bands ranging from 450nm to 650nm.

During training, to simulate the measurements, we first randomly extract $96 \times 96 \times 28$ patches from the training dataset as training labels (ground truth HSI) and randomly extract $96 \times 96$ patches from the {\em real mask} to generate the modulated data. Then the modulated data is shifted in spatial at an interval of two pixels. The spectral dimension of the shifted data is summed up to generate the 2D measurements of size $96 \times 150$ as the network inputs. We use Random flipping and rotation for data argumentation. The peak-signal-to-noise (PSNR) and the structural similarity index (SSIM)~\cite{wang2004image} are both employed to evaluate the performance of the HSI reconstruction methods.

\begin{table*}[htbh]
\caption{The PSNR in dB (left entry in each cell) and SSIM (right entry in each cell) results of the test methods on 10 scenes.}
\centering{
\resizebox{\textwidth}{!}{
\begin{tabular}{|c|c|c|c|c|c|c|c|c|}
\hline
Method  & TwIST~\cite{bioucas2007new} & GAP-TV~\cite{yuan2016generalized}  & DeSCI~\cite{liu2018rank}  & $\lambda$-net~\cite{miao2019lambda} & HSSP~\cite{wang2019hyperspectral}  & DNU~\cite{wang2020dnu} & TSA-Net~\cite{Meng2020TSA-Net}  & Ours                   \\ \hline
Scene1  & 25.16, 0.6996 & 26.82, 0.7544 & 27.13, 0.7479 & 30.10, 0.8492              & 31.48, 0.8577 & 31.72, 0.8634 & 32.03, 0.8920          & \textbf{33.26, 0.9152} \\ \hline
Scene2  & 23.02, 0.6038 & 22.89, 0.6103 & 23.04, 0.6198 & 28.49, 0.8054              & 31.09, 0.8422 & 31.13, 0.8464 & 31.00, 0.8583          & \textbf{32.09, 0.8977} \\ \hline
Scene3  & 21.40, 0.7105 & 26.31, 0.8024 & 26.62, 0.8182 & 27.73, 0.8696              & 28.96, 0.8231 & 29.99, 0.8447 & 32.25, 0.9145          & \textbf{33.06, 0.9251} \\ \hline
Scene4  & 30.19, 0.8508 & 30.65, 0.8522 & 34.96, 0.8966 & 37.01, 0.9338              & 34.56, 0.9018 & 35.34, 0.9084 & 39.19, 0.9528          & \textbf{40.54, 0.9636} \\ \hline
Scene5  & 21.41, 0.6351 & 23.64, 0.7033 & 23.94, 0.7057 & 26.19, 0.8166              & 28.53, 0.8084 & 29.03, 0.8326 & \textbf{29.39, 0.8835} & 28.86, 0.8820          \\ \hline
Scene6  & 20.95, 0.6435 & 21.85, 0.6625 & 22.38, 0.6834 & 28.64, 0.8527              & 30.83, 0.8766 & 30.87, 0.8868 & 31.44, 0.9076          & \textbf{33.08, 0.9372} \\ \hline
Scene7  & 22.20, 0.6427 & 23.76, 0.6881 & 24.45, 0.7433 & 26.47, 0.8062              & 28.71, 0.8236 & 28.99, 0.8386 & 30.32, 0.8782          & \textbf{30.74, 0.8860} \\ \hline
Scene8  & 21.82, 0.6495 & 21.98, 0.6547 & 22.03, 0.6725 & 26.09, 0.8307              & 30.09, 0.8811 & 30.13, 0.8845 & 29.35, 0.8884          & \textbf{31.55, 0.9234} \\ \hline
Scene9  & 22.42, 0.6902 & 22.63, 0.6815 & 24.56, 0.7320 & 27.50, 0.8258              & 30.43, 0.8676 & 31.03, 0.8760 & 30.01, 0.8901          & \textbf{31.66, 0.9110} \\ \hline
Scene10 & 22.67, 0.5687 & 23.10, 0.5839 & 23.59, 0.5874 & 27.13, 0.8163              & 28.78, 0.8416 & 29.14, 0.8494 & 29.59, 0.8740          & \textbf{31.44, 0.9247} \\ \hline
Average & 23.12, 0.6694 & 24.36, 0.6993 & 25.27, 0.7207 & 28.53, 0.8406              & 30.35, 0.8524 & 30.74, 0.8631 & 31.46, 0.8939          & \textbf{32.63, 0.9166} \\ \hline
\end{tabular}}}
\label{tab:Simulate_result}
\vspace{-3mm}
\end{table*}

\begin{figure*}[htbh]
\centering
\includegraphics[width=1.0\linewidth]{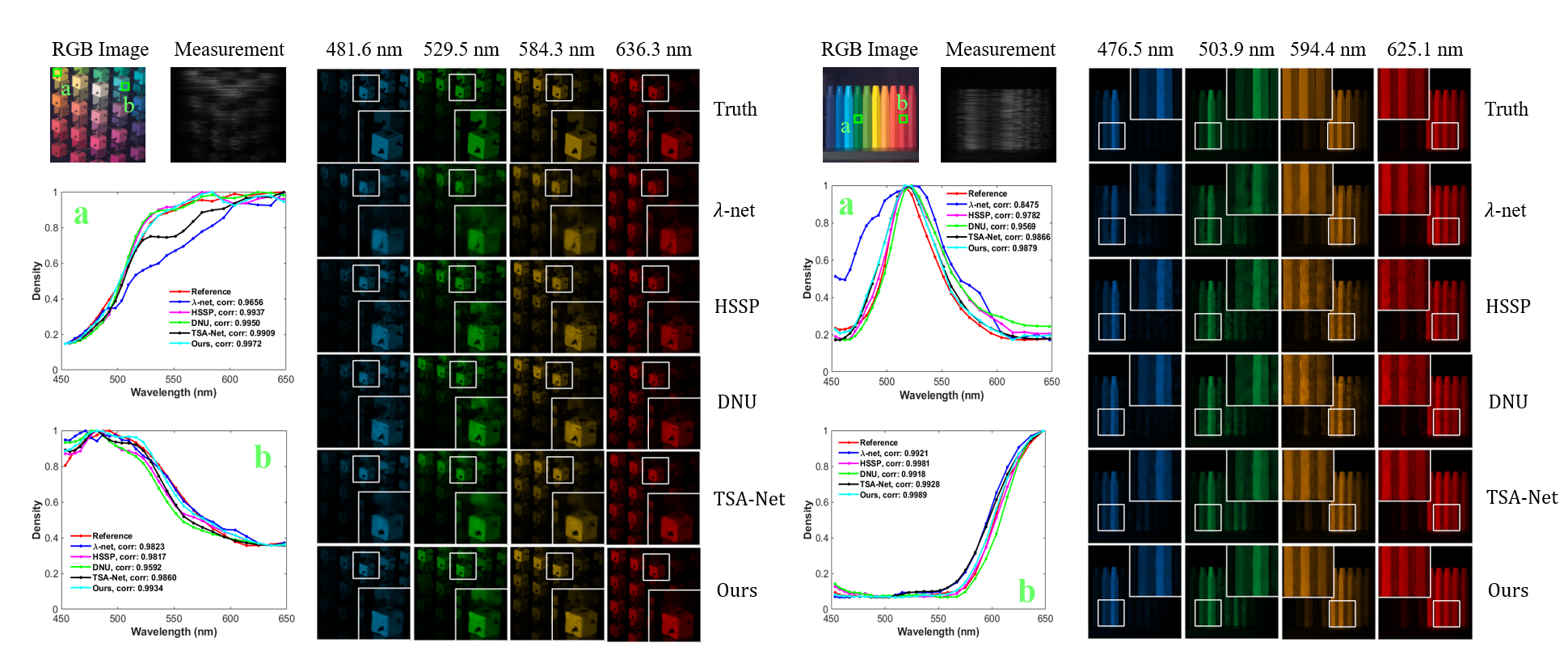}
\vspace{-5mm}
\caption{Reconstructed images of \emph{Scene 2} (left) and \emph{Scene 9} (right) with 4 out of 28 spectral channels by the five deep learning-based methods. Two regions in each scene are selected for analysing the spectra of the reconstructed results. Zoom in for better view.}
\label{fig:Simulate_results}
\end{figure*}

\subsection{Comparison with State-of-the-Art Methods}

We compare the proposed HSI reconstruction method with several state-of-the-art methods, including three model-based methods (i.e., TwIST~\cite{bioucas2007new}, GAP-TV~\cite{yuan2016generalized} and DeSCI~\cite{liu2018rank}) and four deep learning based methods (i.e., $\lambda$-net~\cite{miao2019lambda}, HSSP~\cite{wang2019hyperspectral}, DNU~\cite{wang2020dnu} and TSA-Net~\cite{Meng2020TSA-Net}). As the source codes are unavailable, we re-implemented HSSP and DNU by ourselves. For other competing methods, we use the source codes released by their authors. For the sake of fair comparison, all deep learning methods were {\em re-trained on the same training dataset}. Table \ref{tab:Simulate_result} shows the reconstruction results of these testing methods on the 10 scenes, where we can see that the deep learning-based methods outperform the model-based methods. The proposed method outperforms other deep learning-based methods by a large margin. Specifically, our method outperforms the second best method TSA-Net by 1.17dB in average PSNR and 0.0227 in average SSIM.
Compared with the two deep unfolding methods HSSP and DNU, the improvements by the proposed method over HSSP \cite{wang2019hyperspectral} and DNU \cite{wang2020dnu} are 2.28 dB and 1.89 dB in average, respectively. The HSSP and DNU methods also tried to learn the spatial-spectral correlations of HSIs by two sub-networks without emphasizing image edges and textures. By contrast, we propose to learn the spatial-spectral prior of HSIs by the spatially-adaptive GSM models characterized by the learned local means and variances. The learned GSM models have advantages in adapting to various HSI edges and textures. Fig. \ref{fig:Simulate_results} plots selected frames and spectral curves of the reconstructed HSIs by the five deep learning-based methods. We can see that the HSIs reconstructed by the proposed method have more edge details and less undesirable visual artifacts than the other methods. The RGB images of the 10 scenes and more visual comparison results are shown in the supplementary material (SM).


\begin{figure*}[htbh]
\centering
\includegraphics[width=0.95\linewidth]{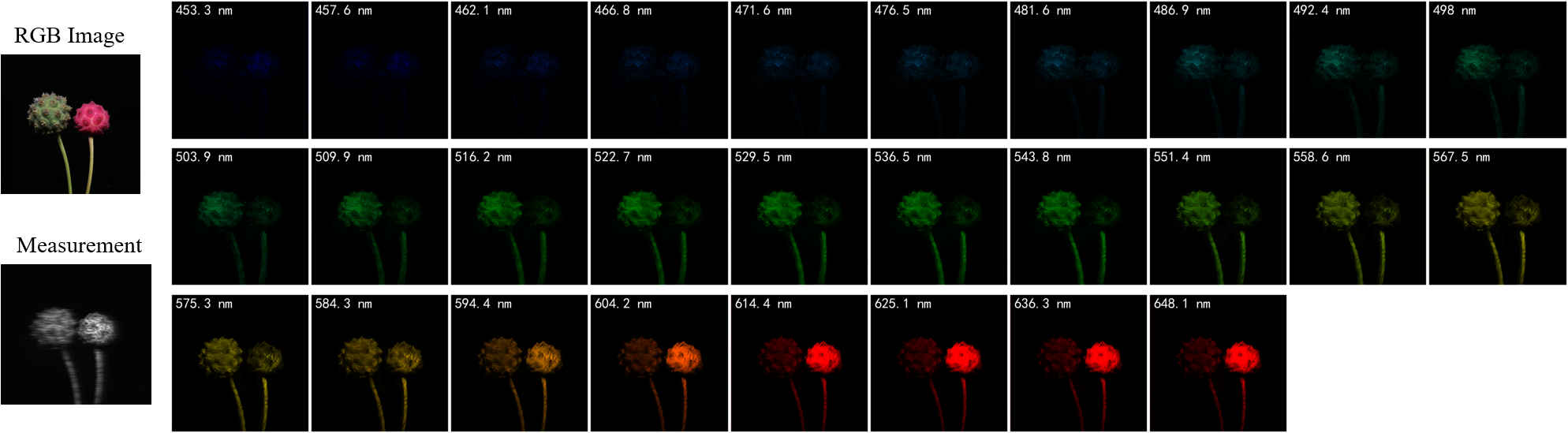}
\caption{Reconstructed images of the real scene (\emph{Scene 4}) with 28 spectral channels by the proposed method. Zoom in for better view.}
\vspace{-2mm}
\label{fig:Real_data_results2}
\end{figure*}

\begin{table}[htbh]
\caption{The average PSNR (left) and SSIM (right) results with five masks by the competing methods.}
\centering{
\resizebox{0.95\columnwidth}{!}{
\begin{tabular}{|c|c|c|c|}
\hline
Method & DNU~\cite{wang2020dnu}  & TSA-Net~\cite{Meng2020TSA-Net}       & Ours                   \\ \hline
mask1  & 30.29, 0.8588              &   30.96, 0.8804            & \textbf{31.38, 0.8979}          \\ \hline
mask2  & 30.46, 0.8516              &   31.23, 0.8875            & \textbf{31.73, 0.9034}          \\ \hline
mask3  & 30.80, 0.8663              &   31.43, 0.8904            & \textbf{31.81, 0.9055}          \\ \hline
mask4  & 30.65, 0.8610              &   31.15, 0.8863            & \textbf{31.58, 0.9038}          \\ \hline
mask5  & 30.74, 0.8631              &   31.46, 0.8939            & \textbf{31.70, 0.9018} \\ \hline
\end{tabular}}}
\vspace{-4mm}
\label{tab:Multiple masks}
\end{table}

\subsection{Multiple Mask Results}

As mentioned before, our proposed network is robust to mask due to the learning of $\tA$ and $\tA^{\top}$.
To verify this, we conducted experiments on compound training and testing datasets that were simulated by applying 5 different masks. The 5 masks of size $256 \times 256$ were extracted at the four corners and the center of the real captured mask \cite{Meng2020TSA-Net}. We only trained {\em a single model} by the proposed network on the compound training dataset to deal with multiple masks, whereas we trained {\em five different models} associated with each mask by the DNU \cite{wang2020dnu} and TSA-Net \cite{Meng2020TSA-Net} methods on the datasets generated by the corresponding mask, respectively. Table \ref{tab:Multiple masks} shows the average PSNR and SSIM results by these testing methods on the 10 scenes. We can see that the proposed method (only trained once on the compound training dataset) still outperforms the other two competing methods that were trained specifically for each mask, verifying the advantages of learning the measurement matrix $\tA$ and $\tA^{\top}$.



\begin{figure}[htbh]
\centering
\vspace{-1mm}
\includegraphics[width=0.9\linewidth]{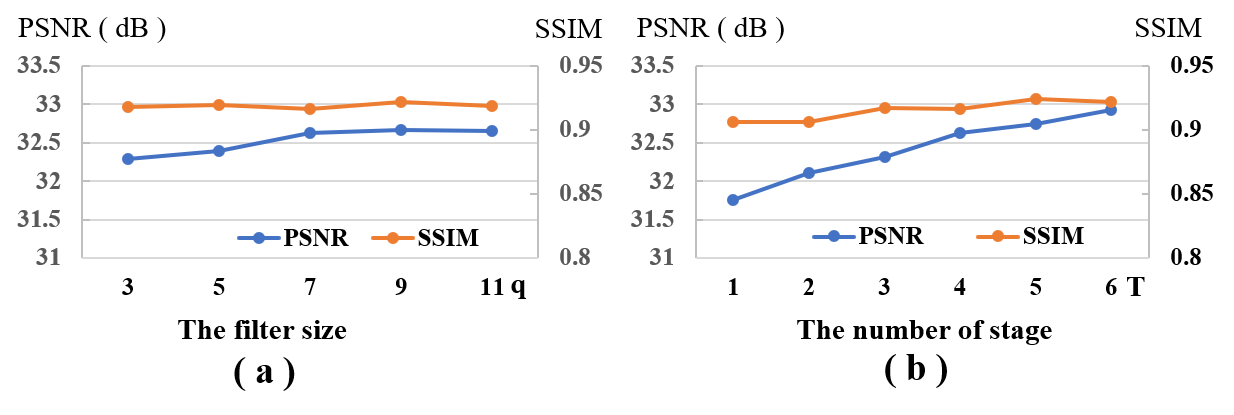}
\vspace{-2mm}
\caption{Ablation study on the effects of (a) the filter size; (b) the number of stage.}
\vspace{-4mm}
\label{fig:Ablation studies}
\end{figure}

\subsection{Ablation Study}

We conduct several ablation studies to verify the impacts of different modules of the proposed network, including the choices of the filter sizes, number of stages and the use of dense connections.

Fig. \ref{fig:Ablation studies} (a) shows the results with different filter sizes, where we can see that larger filter size can improve the HSI reconstruction quality. The improvement flattens out after $q=7$ and thus we set $q=7$ in our implementation. The results with different number of stages are shown in Fig. \ref{fig:Ablation studies} (b), from which we observe that increasing the stage number $T$ leads to better performance. We set $T=4$ in our implementation for achieving a good trade-off between reconstruction performance and computational complexity. We have also conducted a comparison between the proposed network without and with dense connections. The comparison demonstrates that using dense connections can boost PSNR from 30.52dB to 32.63dB and SSIM from 0.8802 to 0.9166.

\section{Real Data Results}
\begin{figure}[htbh]
\centering
\includegraphics[width=0.95\linewidth]{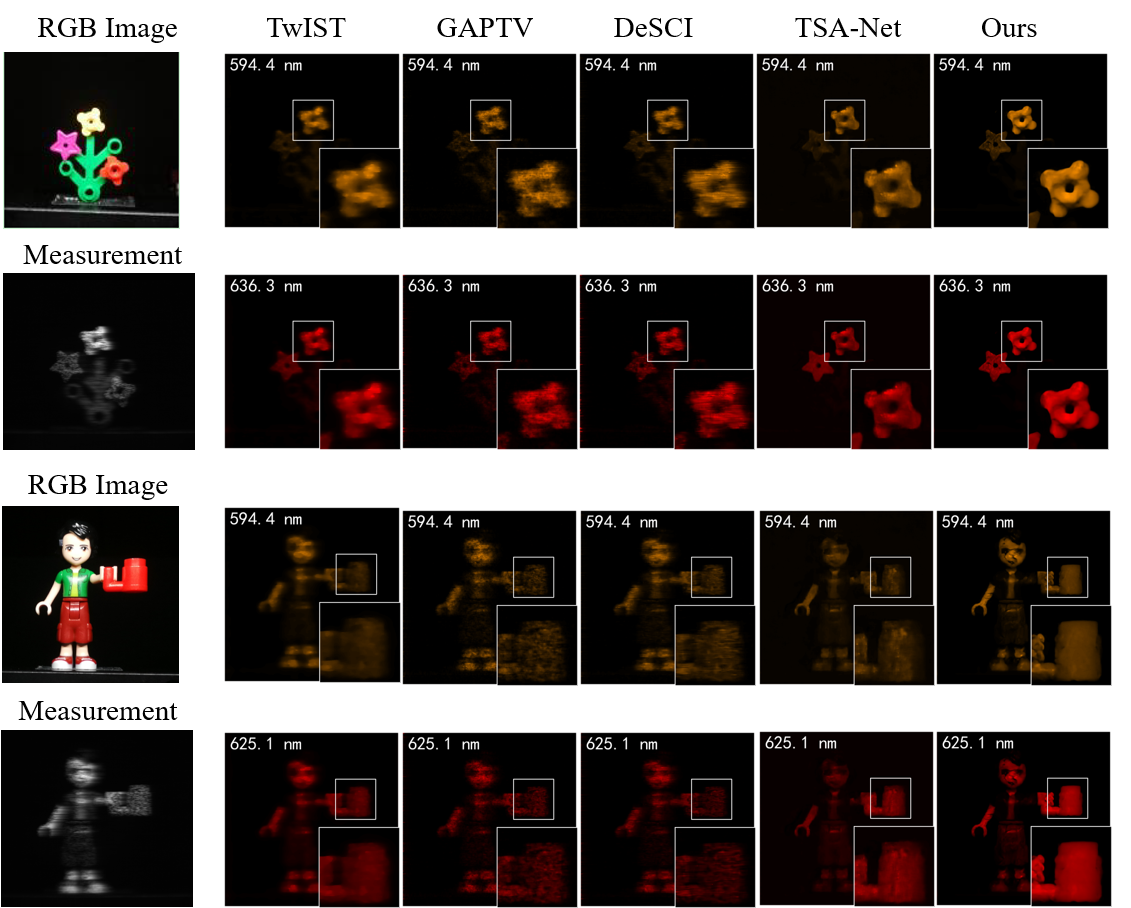}
\caption{Reconstructed images of two real scenes (\emph{Scene 1} and \emph{Scene 3}) with 2 out of 28 spectral channels by the competing methods. Zoom in for better view.}
\vspace{-2mm}
\label{fig:Real_data_results1}
\end{figure}

We now apply the proposed method on the real SD-CASSI system \cite{Meng2020TSA-Net} which captures the real scenes with 28 wavelengths ranging from 450nm to 650nm and has 54-pixel dispersion in the column dimension. Thus, the measurements captured by the real system have a spatial size of $660 \times 714$. Similar to TSA-Net~\cite{Meng2020TSA-Net}, we re-trained the proposed method on all scenes of CAVE dataset and KAIST dataset. To simulate the real measurements, we injected 11-bit shot noise during training. We compare the proposed method with TwIST~\cite{bioucas2007new}, GAP-TV~\cite{yuan2016generalized}, DeSCI~\cite{liu2018rank} and TSA-Net~\cite{Meng2020TSA-Net}. Visual comparison results of the competing methods are shown in Fig. \ref{fig:Real_data_results1}. It can be observed that the proposed method can recover more details of the textures and suppress more noise. Fig. \ref{fig:Measurement&Result} and \ref{fig:Real_data_results2} show reconstructed images of two real scenes (\emph{Scene 3} and \emph{Scene 4}) with 28 spectral channels by the proposed method. More visual results are shown in the SM.

\section{Conclusions}

We have proposed an interpretable hyperspectral image reconstruction method for coded aperture snapshot spectral imaging.
Different from existing works, our network is inspired by the Gaussian scale mixture prior.
Specifically, the desired hyperspectral images were characterized by the GSM models and then the reconstruction problem was formulated as a MAP estimation problem.
Instead of using a manually designed prior, we have proposed to learn the scale prior of GSM by a DCNN.
Furthermore, motivated by the auto-regressive model, the means of the GSM models have been estimated as a weighted average of the spatial-spectral neighboring pixels, and these filter coefficients are estimated by a DCNN as well aiming to learn sufficient spatial-spectral correlations of HSIs.
Extensive experimental results on both synthetic and real datasets demonstrate that the proposed method outperforms existing state-of-the-art algorithms.

Our proposed network is not limited to the spectral compressive imaging such as CASSI and similar systems~\cite{Yuan15JSTSP,Ma2021_LeSTI_OE}, it can also be used in the video snapshot compressive imaging systems~\cite{Qiao2020_CACTI,Qiao2020_APLP,Qiao2021_MicroCACTI,Yuan20PnPSCI}.
Our work is paving the way of real applications of snapshot compressive imaging~\cite{Lu20SEC,Yuan2021_SPM}.

\paragraph{Acknowledgments.} This work was supported in part by the National Key R\&D Program of China under Grant 2018AAA0101400 and the Natural Science Foundation of China under Grant 61991451, Grant 61632019, Grant 61621005, and Grant 61836008.


\clearpage
\newpage

{\small
\bibliographystyle{ieee_fullname}
\bibliography{egbib}
}

\end{document}